\documentclass[12pt]{article}
\usepackage{a4,graphicx,psfrag}
\begin{document}
\makeatletter
\def\fmslash{\@ifnextchar[{\fmsl@sh}{\fmsl@sh[0mu]}}
\def\fmsl@sh[#1]#2{%
  \mathchoice
    {\@fmsl@sh\displaystyle{#1}{#2}}%
    {\@fmsl@sh\textstyle{#1}{#2}}%
    {\@fmsl@sh\scriptstyle{#1}{#2}}%
    {\@fmsl@sh\scriptscriptstyle{#1}{#2}}}
\def\@fmsl@sh#1#2#3{\m@th\ooalign{$\hfil#1\mkern#2/\hfil$\crcr$#1#3$}}
\makeatother
%
\thispagestyle{empty}
\begin{titlepage}

\begin{flushright}
TTP 00--21 \\
September 22, 2000
\end{flushright}

\vspace{0.3cm}
\boldmath
\begin{center}
\Large\bf  Radiatively corrected shape function  \\ 
           for inclusive heavy hadron decays
\end{center}
\unboldmath
\vspace{0.8cm}

\renewcommand{\thefootnote}{\fnsymbol{footnote}}
\begin{center}
{\large Thomas Mannel\footnote[2]{From Oct.\ 1st 2000: Theory Group,
CERN, CH--1211 Geneva 23}}
and {\large Stefan Recksiegel\footnote[4]{
From Oct.\ 1st 2000: Theory Group, KEK, Tsukuba, Ibaraki, 305-0801, Japan}}
\\[1mm]
{\sl Institut f\"{u}r Theoretische Teilchenphysik,
     D--76128 Karlsruhe, Germany.} 
\end{center}

\vspace{\fill}

\begin{abstract}
\noindent
We discuss the non-perturbative and the radiative corrections to 
inclusive $B$ decays from the point of view known from QED
corrections to high energy $e^+ e^-$ processes. Here the 
leading contributions can be implemented through the so called
``radiator function'' which corresponds to the shape function 
known in heavy hadron decays. In this way some new insight 
into the origin of the shape function is obtained. As a 
byproduct, a parameterization of the radiatively corrected shape
function is suggested which can be implemented in Monte Carlo studies
of inclusive heavy hadron decays.  
\end{abstract}
\end{titlepage}

%
%
%
\section{Introduction}
It has become a standard method to describe inclusive heavy hadron 
decays in the framework of an expansion in inverse powers of the heavy
quark mass $m_Q$. This approach allows us to compute lifetimes and 
branching ratios in a QCD framework, where radiative corrections due 
to gluon radiation can be systematically computed, while 
non--perturbative effects can be parameterized to obtain
estimates on their size. 

This method also yields reliable answers for differential decay rates 
such as energy spectra or invariant mass spectra that hold in 
practically all phase space with the exception of certain endpoint 
regions.
These regions are characterized by the kinematic situation in which the 
heavy quark decays into a practically massless light quark, and all the 
emitted gluons are collinear to the light quark. In such a situation the 
hadronic invariant mass in the final state is small while the hadronic 
energy is still large. 

It is well known that in these regions the standard $1/m_Q$ expansion 
breaks down and has to be replaced by an expansion in twists, very similar 
to what happens in deep inelastic scattering. The leading twist term has 
been discussed extensively \cite{NeubertBigi} and is described in terms of a single
non--perturbative function, the light cone distribution function of the 
heavy quark. It has been shown that this approach is in for a certain
range of parameters equivalent to the ACCMM model, which is very popular
in the phenomenological description of inclusive heavy flavour decays.

However, it is by not trivial to combine the radiative QCD corrections 
with this light cone distribution function in a consistent way.
It has been observed in 
\cite{Korchemsky:1994jb} that one may think of the light cone distribution
function as defined in \cite{NeubertBigi}
as a convolution of a ``soft'' 
and a ``jet-like'' contribution, where the latter contains all the collinear 
singularities such as the Sudakov Logarithms. 

In the present note we give a practical parameterization of the radiatively
corrected light cone distribution function 
which can be used to 
implement the leading radiative and non-perturbative corrections into a 
decay rate known at tree level. The spirit of this approach is very much 
along the lines which have been used in high energy $e^+ e^-$ or $e p$
processes to implement the leading radiative corrections, e.g.\ for the 
$Z_0$ line shape \cite{Nicrosini:1989ax}. 
In these processes the the leading logarithmic  resummation  
of the QED radiative corrections leads to a so called ``radiator function'' 
which is universal and has to be convoluted with the tree level process
to obtain a radiatively corrected cross section. However, in the case of 
$B$ decays the situation is a little more complicated due to the presence 
of non--perturbative effects and the Sudakov logarithms, but the physics
is essentially the same. 

In the next section we discuss the formalism and introduce 
the parameterization for the analogue of the radiator
function for heavy hadron decays. In section 3 we consider a few applications 
of our formalism, finally we conclude. 

\section{``Radiator Functions'' in heavy hadron decays}
\label{RadiatorFunctions}
%
%
%
The situation in inclusive heavy hadron decays is completely analogous 
to the one in high energy $e^+ e^-$--collisions. In both cases ``initial 
state radiation'' diminishes the cms energy of the ``partonic'' process. 
While in $e^+ e^-$ collisions the initial state radiation is the emission 
of soft and collinear (to the incident particles) photons, 
it is the emission of soft and collinear gluons in heavy hadron 
decays. The soft gluons correspond to a completely 
non--perturbative function, while the collinear ones lead to Sudakov--like
Logarithms, for which also resummation formulae exist. 

In high energy $e^+ e^-$ processes the bulk of the QED radiative corrections
is obtained by a convolution of the ``partonic'' (i.e.\ tree level) cross
section with a radiator function, such that 
\begin{equation} \label{ee}
d \sigma (s) = \int\limits_0^1 dx_1\,dx_2\,\,
D_{e^-}(x_1,s,m_e^2) D_{e^+}(x_2,s,m_e^2)\,
d \sigma_{tree} (x_1 x_2 s)  
\end{equation}
where the tree-level cross section is taken at
a reduced cms energy $x_1 x_2 s$.  
The radiator function $D_{e^{-(+)}}(x)$ 
\begin{equation} \label{radiator}
D(x,s,m_e^2) = \delta(1-x)
+\int\limits_{m_e^2}^s {dk^2\over k^2}{\alpha(k^2)\over 2\pi} P(x)
 + \dots \approx
   \delta(1-x) + {\alpha\over 2\pi}
                 \ln\left(\frac{s}{m_e^2}\right) P(x) + \cdots
\end{equation}
with
\begin{equation}
P(x) = \left({1+x^2\over 1-x}\right)_+ =  
{1+x^2\over 1-x}-
\delta(1-x)+\lim_{\epsilon \to 0} \int_\epsilon^1\! dz{1+z^2\over 1-z}
\end{equation}
describes the energy loss of the 
incident particles due to radiation of soft and collinear photons
including leading logarithms of the form $\ln (s/m_e^2)$. The ellipses
denote terms of higher order, which we shall not discuss here. 
Furthermore, the radiator function depends on the scale of the
process $\sqrt{s}$ and on a 
``regularization scale'' $m_e$ cutting off the collinear singularities
of the photon emission. 

While the full radiative corrections (i.e. the ones beyond the leading logs)
have to be discussed at the level of amplitudes, the leading logarithmic 
radiative corrections are described on the level of probabilities: $D(x)$ 
can be interpreted as the probability that the incident particles loose 
the energy fraction $x$ due to radiation. 

It is well known that this concept is universal and can be applied to
the QCD case as well. In deep inelastic scattering the radiator functions
become the parton distributions and the scale corresponding to the lower
limit of the integration in (\ref{radiator}) becomes the factorization
scale. 

In inclusive heavy hadron decays this works completely
analogous. In this case the radiator function is the shape function
or the light cone distribution function
as given in \cite{NeubertBigi}. The physical picture is the same:
The cms--energy available for the decay is the mass $M$ of the heavy hadron, 
but the partonic decay happens at a smaller energy $m_Q + k_+$, where 
now $- m_Q \le k_+ \le \bar\Lambda$ corresponds to the scaling variable
$x$ of the radiator function in (\ref{radiator}). 

The masses of the hadron $M$ and of the mass $m_Q$ of the heavy quark
are related to order $1/m_Q$ by
\begin{equation} \label{mass}
M = m_Q + \bar\Lambda  
\end{equation}
The corresponding decay rates in the infinite mass limit are in analogy to 
(\ref{ee}) \cite{NeubertBigi}  
\begin{equation} \label{convol}
d\Gamma = \int\limits_{-\infty}^{\bar\Lambda} dk_+ \, R(k_+,m_Q) \, 
          d\Gamma_{tree} (m_Q + k_+) + {\cal O}(1/m_Q^2)  
\end{equation}
where $d\Gamma_{tree}$ is the tree-level partonic rate and $R(k_+,\mu)$
is the ``radiator function'' for heavy hadron decays. 

It has been argued in \cite{Korchemsky:1994jb} that this function can 
again be decomposed into a ``soft'' and a ``jet-like'' part, such that 
\begin{equation} \label{radcorr}
R(k_+, \mu) = 
\int\limits_{-\infty}^{\bar\Lambda} dl_+ \, J(k_+ - l_+,\mu) \, f(l_+,\mu) 
\end{equation}
where $J$ can be computed in perturbation theory and $f$ is a
non-perturbative quantity. The factorization into a perturbative
part and a non-perturbative contribution involves a factorization
scale $\mu$, the dependence on which has been considered in
\cite{BalzereitKilianMannel, Aglietti}.  

It is interesting to consider the dependence of this formalism on the
on the heavy quark mass; we shall use this information later to adjust
the parameters. It is well known that the total rates are
proportional to the mass of the heavy quark $m_Q$ such that
\begin{equation} \label{totalrate}
\Gamma = \gamma m_Q^5 \left(1 + b \frac{\alpha_s (m_b)}{\pi} + \cdots 
         + {\cal O}(1/m_Q^2)   \right)
\end{equation}
where $b$ is a constant depending on the process under consideration
and also on the choice of the mass definition. In fact, using the pole mass
in (\ref{totalrate}) generically yields large and negative coefficients
$b$ and it has been argued \cite{Bigietal} that a different choice of the 
mass definition renders $b$ small for all processes. For instance, using 
instead of the pole mass the $\overline{MS}$ mass and employing the 
one loop relation  
\begin{equation} \label{msbarmass} 
m_b^{pole} = m_b^{\overline{MS}}(\mu) \left(1 +
\frac{\alpha_s}{\pi} \left[\frac{4}{3} - 
                     \ln\left(\frac{m_b^2}{\mu^2}\right) \right] \right)
\end{equation}
the corresponding coefficient for the first-order correction to the 
rate becomes 
\begin{equation}
b \to b +\frac{20}{3} - 5 \ln\left(\frac{m_b^2}{\mu^2}\right)
\end{equation}  

On the other hand, we may also compute the total rate in the shape function
formalism by integrating (\ref{convol}). Inserting the tree level result   
for the total rate, we have
\begin{equation} 
\Gamma = \int\limits_{-\infty}^{\bar\Lambda} dk_+ \, R(k_+,m_Q) \, 
         \gamma m_Q^5 \left(1 + 5 \frac{k_+}{m_Q}
         + {\cal O}(1/m_Q^2)   \right)
\end{equation}
We use the conditions on the shape function
\begin{eqnarray} \label{norm1}
\int\limits_{-\infty}^{\bar\Lambda} dk_+ f(k_+ ,\mu) &=& 1 \\
\label{norm2}
\int\limits_{-\infty}^{\bar\Lambda} dk_+ k_+ f(k_+ ,\mu) &=& \delta m_Q (\mu)
\end{eqnarray}
Note that   
the second relation is the equation of motion for the heavy
quark, where a possible residual mass term $\delta m_Q$ \cite{FalkNeubert},
which represents the difference between the mass used in the Lagrangian
and the mass used in the definition of the heavy quark momentum $p = m_Q v$,
has been kept. Obviously, at tree level we recover the total rate 
$\Gamma = \gamma m_Q^5$, if we have $\delta m_Q =0 $ 

Our aim is to include the large radiative correction coming from the
quark mass into a radiatively corrected shape function as defined in
(\ref{radcorr}). This involves the first moment of $R$ defined in
(\ref{radcorr}) and we obtain
\begin{equation} 
\Gamma = \gamma m_Q^5 \left[\left(
\int\limits_{-\infty}^{\bar\Lambda} dk_+ \, R(k_+,m_Q) \right) 
+ 5 \left(
\int\limits_{-\infty}^{\bar\Lambda} dk_+ \frac{k_+}{m_Q} R(k_+,m_Q)
    \right) + {\cal O}(1/m_Q^2) \right]   
\end{equation}
We assume that the non-perturbative function $f$ satisfies the normalization
conditions (\ref{norm1},\ref{norm2}) at some small scale $\mu$, and  
we find
\begin{equation} 
\Gamma = \gamma m_Q^5 \left[ \left(
\int\limits_{-\infty}^0 dk_+ \, J(k_+,\mu) \right) 
 \left(1+ 5 \frac{\delta m_Q}{m_Q} \right) + 
\int\limits_{-\infty}^0 dk_+ \frac{k_+}{m_Q} J(k_+,\mu) 
         + {\cal O}(1/m_Q^2)    \right] 
\end{equation}
This shows that the radiative corrections to the total rate
are mainly given by the integral over the ``jet-like'' function.  
This is a universal function and hence we again find that some large portion 
of the radiative corrections (namely the leading log contribution) is 
universal.

However, there also appears a contribution of order $1/m_Q$ which 
according to (\ref{totalrate}) has to be zero. This means that the
residual mass term has to be chosen in such a way that the corrections
of order $1/m_Q$ vanish, i.e.
\begin{equation}
\delta m_Q = - \int\limits_{-\infty}^0 dk_+ \, k_+ J(k_+,\mu) \,\,\,/
\int\limits_{-\infty}^0 dk_+ \, J(k_+,\mu)
\end{equation}

Let us now discuss the ``jet-like'' function explicitely up to order 
$\alpha_s$. The leading logarithmic piece is given by
\begin{equation} \label{jet1}
J (k_+, \mu) = \left( 1 + \frac{\alpha_s}{3\pi} b_1  \right) \delta(k_+) 
- \frac{4\alpha_s}{3\pi} \Theta(-k_+) 
  \left(\frac{\ln (-k_+/\mu)}{-k_+} \right)_+
\end{equation}
where the double logs $\ln (-k_+/m_Q)/k_+$ are universal, while the single
logs are not universal; their size is fixed by an appropriate choice of the 
scale $\mu$. 

Using (\ref{jet1}) we can compute the normalization and the first moment
for which we get
\begin{eqnarray}
\int\limits_{-\infty}^0 dk_+ J(k_+,\mu) &=& 1 + \frac{\alpha_s}{3\pi} b_1 \\
\int\limits_{-\infty}^0 dk_+ k_+ J(k_+,\mu) &=& 
\frac{4 \alpha_s}{3 \pi} M \left[\ln\left(\frac{M}{\mu}\right) - 1 \right] 
\end{eqnarray}
where we have inserted $-M$ as a lower cut-off for the $\xi_+$ integral.  

This cut off dependence reflects the fact that although all moments of
the non-perturbative function $f$ exist\footnote{This means that $f$ has to
         decrease exponentially as $k_+ \to -\infty$}, the moments of the
radiatively corrected function $R$ pick up power divergencies. The
physical origin of this problem is the ``bremsstrahlung'' spectrum of
the gluon emission behaving as $1/(-k_+)$ and thus the the first moment
picks up a linear divergence.

Clearly the physical limit of $k_+$ is $-m_Q$ and thus we shall identify
$M= m_Q$. The absence of $1/m_Q$ corrections in the total rate thus means
that the first moment of the non-perturbative function has to be
\begin{equation}
\delta m_Q = - \frac{4 \alpha_s}{3 \pi} m_Q
             \left[\ln\left(\frac{m_Q}{\mu}\right) - 1 \right]
\end{equation}
while $b_1$ is given by the large and universal correction to the total rate.

Again we consider using a different mass definition in the shape function 
formalism. For instance, using the $\overline{MS}$ mass defined in 
(\ref{msbarmass}) we get for the mass difference
\begin{equation}
m_b^{pole} -m_b^{\overline{MS}} (\mu) = 
                      m_b \frac{\alpha_s}{\pi} \left[\frac{4}{3} - 
                     \ln\left(\frac{m_b^2}{\mu^2}\right) \right] 
\end{equation}
which at least for the non-logarithmic terms agrees with $\delta m_Q$ computed 
from the first moment of the shape function. This leads to the conjecture, 
that the bulk of the radiative corrections at the leading twist level can 
indeed be absorbed by an appropriate choice of the mass definition, even for
differential quantities affected by the shape function. The shape function 
itself depends on the mass definition via its first moment, which has to be 
shifted by a $\delta m_Q$ corresponding to the difference between the 
mass definition employed and the pole mass.   

For practical applications, (\ref{jet1}) is still not acceptable, since the
``jet-like'' function is interpreted in a similar way as in the case of
$e^+e^-$ annihilation as a probability of having the tree level process at
a reduced cms-energy. However, (\ref{jet1}) is not positive definite.
This problem has been discussed in \cite{MannelRecksiegel2} where it was
shown that sub-leading contributions render the decay rates positive. 
Thus for phenomenological applications we
add terms corresponding to sub-leading pieces. Of course,
in general terms of sub-leading twist depend on the specific process,
but here we add universal terms which are fixed by comparing to
the results of \cite{MannelRecksiegel2}. The formula we are going to use
is
\begin{equation}
J (k_+, \mu) = \left( 1 + \frac{\alpha_s}{3\pi} b_1 \right)  \delta(k_+) 
+ \frac{\alpha_s}{3\pi} \Theta(-k_+) \left[ -4 
     \left(\frac{\ln (-k_+/\mu)}{-k_+} \right)_+ + b_2 \frac{k_+}{m_Q^2} 
      + b_3 \frac{1}{m_Q} \right] \label{J}
\end{equation}
The three constants appearing in (\ref{J}) have to be determined from
some input. One condition is that the large radiative corrections
are taken into account, which fixes $b_1$ as discussed above.
Furthermore, the minimal, kinematically allowed  value of $k_+$ is $-m_Q$ 
where the rate should vanish, thus we require 
\begin{equation}
J ( k_+ = -m_Q, \mu) = 0 \, .
\end{equation}
Furthermore, the lower limit of the $k_+$ integration is now $m_Q$.

These conditions can be fulfilled with suitable $b_1$ and $b_3$
for any choice of $b_2$. The coefficient $b_2$ multiplies a
term of sub-leading twist \cite{MannelRecksiegel2} that could in
principle be neglected; for a vanishing $b_2$, however,
$J (k_+ , \mu)$ becomes negative for some $-m_Q < k_+ < 0$.
Comparison with explicit calculations \cite{MannelRecksiegel2}
suggests a value for $b_2$ which in turn fixes the constant
$b_3$. This way we obtain 
\begin{eqnarray}
J (k_+, \mu) &=& \left( 1 - \frac{\alpha_s}{3\pi} 
          b +8\ln{\mu\over m_Q}-4  \right) \delta(k_+) 
          + \frac{4\alpha_s}{3\pi} \Theta(-k_+) \\ \nonumber 
&& \left[ 
     \left(-\frac{\ln (-k_+/\mu)}{-k_+} \right)_+ 
     + \left(2\ln\frac{\mu}{m_Q}-2\right)\frac{-k_+}{m_Q^2}
     + \frac{-3\ln(\mu/m_Q)+2}{m_Q}
\right]
\end{eqnarray}
Here $b$ is from the ${\cal O}(\alpha_s)$ correction to the total decay rate
(\ref{totalrate}).

The scale $\mu$ is not fixed by these considerations; explicit next-to-leading 
order calculations show that $\mu$ is about $0.1 m_Q ... 0.2 m_Q$ and thus 
is a low scale. Again this is similar to the high-energy $e^+ e^-$ case. 

To parameterize the non-perturbative contributions we use
the model shape function from \cite{MannelNeubert},
\begin{equation}
 f(k_+)={32\over \pi^2 \sigma}\, \left[1-\left(k_+\over 
   \sigma\right)\right]^2 \, \exp\left[-{4\over \pi}
   \left(1-\left(k_+\over\sigma\right)\right)^2 \right] \Theta
   \left[1-\left({k_+\over\sigma}\right)\right] \label{sfunctionmannel}
\end{equation}
This function has vanishing first moment as required at tree level, i.e.
$\delta m_Q = 0$. A non-vanishing first moment as required once radiative
corrections are taken into account is achieved by shifting the argument 
of $f$ by $k_+ \to k_+ + \delta m_Q$, where delta $m_Q$ is calculated from 
the first moment of (\ref{J}). 

\begin{figure}[h]
 \begin{center}
 {\large \psfrag{Xaxis}{$k_+/m_b$} 
  \psfrag{Yaxis}{$R(k_+,\mu$)}
 \includegraphics[width=14cm]{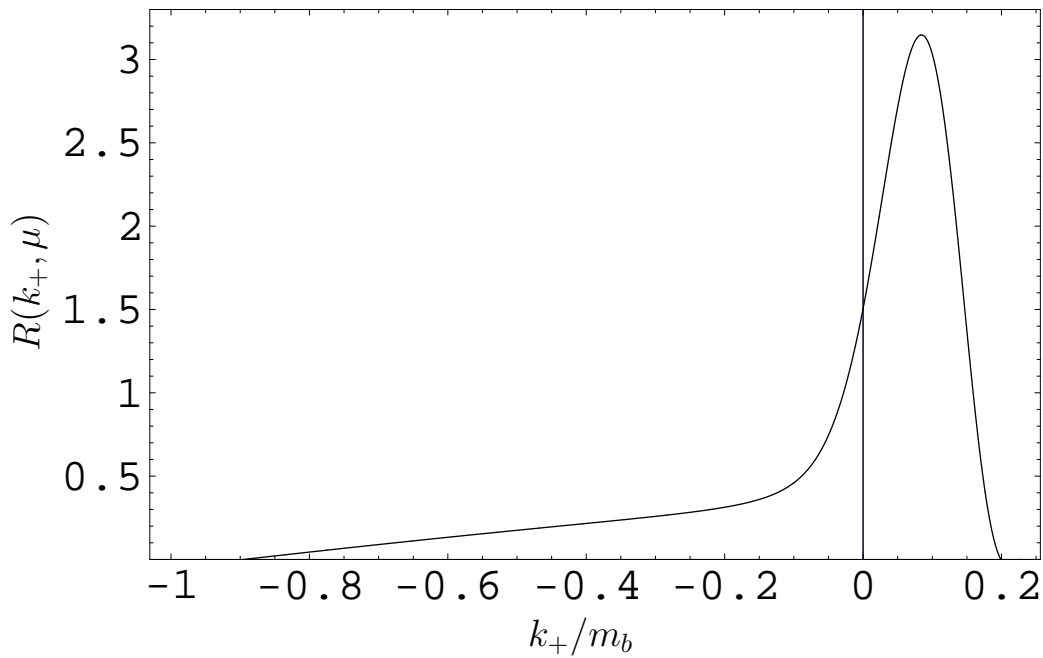}}
 \caption{Radiatively corrected shape function $R(k_+,\mu)$.}
 \label{bsg} 
 \end{center}
\end{figure}

The radiatively corrected shape function $R(k_+,\mu)$
for a typical $\mu=0.1146 m_b$ (as obtained from an explicit
calculation of the ${\cal O}(\alpha_s)$ radiative corrections
to $b\to u\ell\bar\nu_\ell$) is shown in figure~\ref{bsg}.
Since for the decay $B\to X_s\gamma$ the photon energy spectrum is 
(except for factors) just the shape function itself with the endpoint 
shifted to $M/2$, this diagram at the same time shows the spectrum
for $B\to X_s\gamma$.

\begin{figure}[h]
 \begin{center}
 {\large \psfrag{Xaxis}{$E_\ell/m_b$} 
  \psfrag{Yaxis}{$d\Gamma_{B \to X_u \ell \bar{\nu}}/dE_\ell$}
 \includegraphics[width=14cm]{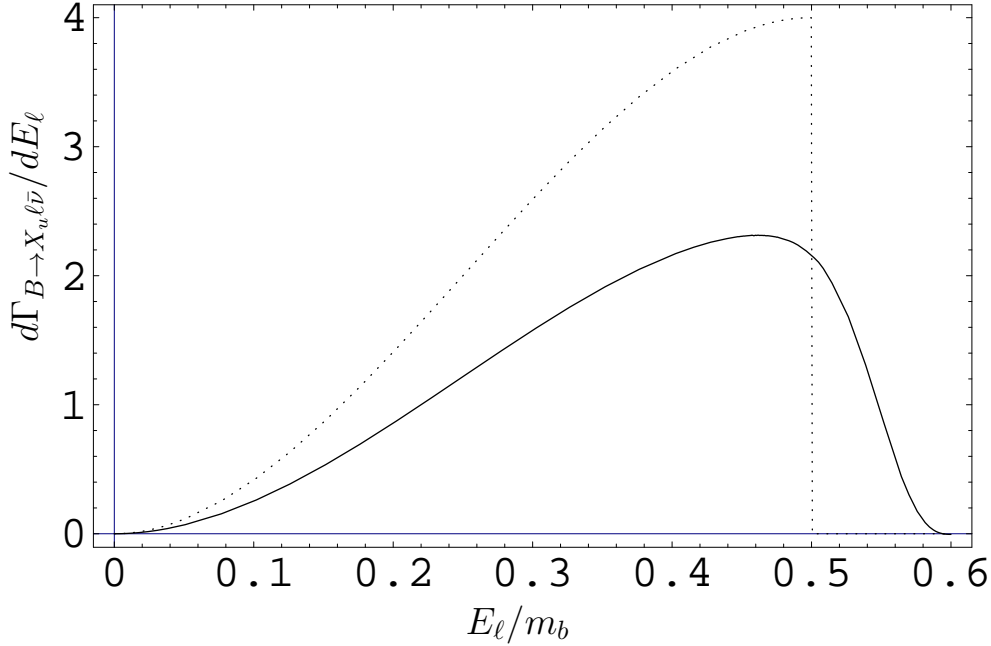}}
 \caption{Lepton energy spectrum for $B \to X_u \ell \bar{\nu}$,
     tree level (dotted line), and including corrections parameterized
     with the radiator function formalism (solid line). The rate
     is given in units of $G_F^2 m_b^5 \left|V_{ub}^2\right| /
     96 \pi^3$}
 \label{bulnu} 
 \end{center}
\end{figure}

In the case of the decay $B\to X_u \ell \bar{\nu}$ the
spectrum is a step function at maximal lepton energy, which is 
smoothened by the shape function effect. The corresponding results are 
shown in figure~\ref{bulnu}

\section{Applications to the decay $B\to X_s \ell^+\ell^-$} 

As an application we employ the radiatively corrected shape function 
for the lepton energy spectra in $B \to X_s \ell^+\ell^-$. While it 
is well known that the shape functions do not play a significant role 
in the leptonic  invariant mass spectrum, they significantly 
modify the energy spectra of the leptons. 

Of particular interest are the lepton energy spectra, where the flavour 
of the decaying $B$ meson has been tagged. These spectra have been suggested 
to test the coefficients $C_9$ and $C_{10}$ of the effective weak Hamiltonian 
governing weak decays of $b$ quarks. In addition, for experimental reasons 
an energy cut has been applied to the second lepton.  

\begin{figure}[h]
 \begin{center}
 {\large \psfrag{Xaxis}{$E_{\ell^+}/m_b$} 
  \psfrag{Yaxis}{$d\Gamma_{B \to X_s \ell^+\ell^-}/dE_{\ell^+}$}
 \includegraphics[width=14cm]{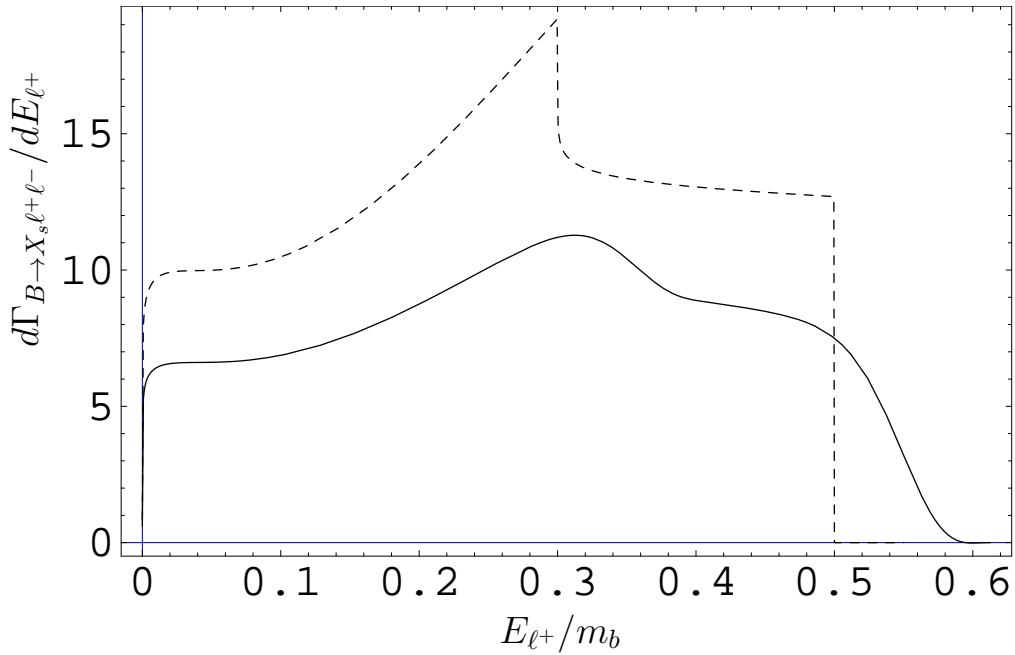}}
 \caption{Spectra for the energy of the positively charged lepton
    in $B \to X_s \ell^+\ell^-$ with (solid) and without (dashed)
    shape function corrections. Both spectra include a cut
    of $0.2m_b$ on the energy of the negatively charged lepton.
    The spectra are given in units of $G_F^2 \alpha^2 \left|
    V_{ts}^*V_{tb}\right|^2 m_b^5/3(2\pi)^5$ 
    Standard values have been used for the Wilson coefficients;
    all our plots use $\alpha_s(m_b)=0.215$, and the shape function 
    parameter is $\sigma/m_b=0.57/4.71$.}
 \label{bslltree} 
 \end{center}
\end{figure}

The partonic spectra show pronounced, step function like structures which 
will be affected strongly by the shape function effects. 
In figure (\ref{bslltree}) we show the 
$E_{\ell^+}$ spectrum with a cut of $0.2m_b$ on the
energy of the negatively charged lepton both with
and without shape function effects.
It can be seen how the radiatively corrected shape function
not only shifts the endpoint of the spectrum and adjusts the
rate but also smears out the features of the
partonic spectrum.

\section{Conclusions}
The question of how to implement non-perturbative contributions due to 
bound state effects in combination with a consistent treatment of radiative 
corrections has been discussed form a new point of view. Comparing the 
situation in $B$ decays to the case of $e^+ e^-$ annihilation we look at 
the shape function as the analogue of the radiator function in high 
energy $e^+ e^-$ processes. In both cases the bulk of the radiative 
corrections (and in $B$ decays the main non-perturbative effects) can 
be accounted for by a convolution of the tree level rate with an appropriate 
function, which can be interpreted as a probability for a certain energy 
loss. Since this works at the level of probabilities, this approach can be 
easily implemented into Monte Carlo programs, which has been done 
quite successfully for the case of high energy $e^+ e^-$ processes
\cite{MonteCarlos}. 

For the case of $B$ decays the situation is complicated by the presence
of non-perturbative effects. These have to be parameterized in terms of a 
so called shape function, which then is combined with the leading radiative 
corrections into the analogue of the radiator function. 

The energy available in the $B$ decay process depends on the mass of the
heavy quark in the initial state. On the other hand, the $b$ quark mass 
is not a physical quantity and thus one has to consider the dependence on 
the mass definition. It turns out that the main part of the radiative
corrections can indeed be absorbed into an appropriate mass definition; 
this fact is well known for the total rates, but it seems to be true also 
for the leading twist contributions parameterized by the shape function. 

We have given a parameterization of the radiatively corrected shape function 
which includes some sub-leading twist contributions which have been fitted in 
such a way that the explicitely known decays of the heavy-to-light type are   
well reproduced. This parameterization may be useful for practical 
applications, in particular for Monte Carlo simulations of $B$ decays. 

\section*{Acknowledgments} 

SR and TM acknowledge the support of the DFG 
Graduiertenkolleg ``Elementarteilchenphysik an Beschleunigern''; 
TM acknowledges the support of the DFG Forschergruppe 
``Quantenfeldtheorie, Computeralgebra und MonteCarlo Simulationen''.


\begin{thebibliography}{99}

\bibitem{NeubertBigi}
M. Neubert, Phys. Rev. {\bf D 49},  3392  (1994);
M. Neubert, Phys. Rev. {\bf D 49},  4623  (1994);
T. Mannel and M. Neubert, Phys. Rev. {\bf D 50},  2037  (1994);
I.~I. Bigi, M.~A. Shifman, N.~G. Uraltsev, 
  and A.~I. Vainshtein, Int. J. Mod. Phys. {\bf A 9},  2467  (1994).

\bibitem{Korchemsky:1994jb}
G.~P. Korchemsky and G. Sterman, Phys. Lett. {\bf B 340},  96  (1994)

\bibitem{Nicrosini:1989ax}
see e.g.\ O.~Nicrosini and L.~Trentadue
in {\it Radiative Corrections to $e^+e^-$ Collisions},
   Edited by J.H.\ K\"uhn, Springer-Verlag, 1989

\bibitem{BalzereitKilianMannel}
C.~Balzereit, T.~Mannel and W.~Kilian,
Phys.\ Rev.\  {\bf D 58}, 114029 (1998)

\bibitem{Aglietti}
U.~Aglietti,
``The shape function in field theory,''
CERN-TH/2000-278, hep-ph/0009214;
U.~Aglietti and G.~Ricciardi,
``The structure function of semi-inclusive heavy flavour decays in field theory,''
CERN-TH-2000-071, DSF-T-7-00, hep-ph/0003146.

\bibitem{Bigietal}
I.~I. Bigi, M.~A. Shifman, N.~G. Uraltsev, 
  and A.~I. Vainshtein in \cite{NeubertBigi}

\bibitem{FalkNeubert}
A.~F.~Falk, M.~Neubert and M.~Luke,
Nucl.\ Phys.\  {\bf B 388}, 363 (1992)

\bibitem{MannelRecksiegel2}
T.~Mannel and S.~Recksiegel,
Phys.\ Rev.\  {\bf D 60} (1999) 114040

\bibitem{MannelNeubert}
T. Mannel and M. Neubert in \cite{NeubertBigi}

\bibitem{MonteCarlos}
see e.g.\ H.~Anlauf, H.~D.~Dahmen, A.~Himmler, P.~Manakos, T.~Mannel and T.~Ohl,
Nucl.\ Phys.\ Proc.\ Suppl.\  {\bf 37 B}, 81 (1994);
S.~Jadach, B.~F.~Ward and Z.~Was,
Nucl.\ Phys.\ Proc.\ Suppl.\  {\bf 89}, 106 (2000)

\end{thebibliography}
\end{document}